# Mapping Morphological and Structural Properties of Lead Halide Perovskites by Scanning Nanofocus XRD

*Samuele Lilliu[1,*], Thomas G. Dane[2], Jonathan Griffin[1], Mejd Alsari[3], Alexander T. Barrows[1], Marcus S. Dahlem[4], Richard H. Friend[3], David G. Lidzey[1], J. Emyr Macdonald[5]*

Scanning nanofocus X-ray diffraction (nXRD) performed at a synchrotron is used for the first time to simultaneously probe the morphology and the structural properties of spin-coated $CH_3NH_3PbI_3$ (MAPI) perovskite films for photovoltaic devices. MAPI films are spin-coated on a $Si/SiO_2$/PEDOT:PSS substrate held at different temperatures during the deposition in order to tune the perovskite film coverage, and then investigated by nXRD, scanning electron microscopy (SEM) and grazing incidence wide angle X-ray scattering (GI-WAXS). The advantages of nXRD over SEM and GI-WAXS are discussed. A method to visualize, selectively isolate, and structurally characterize single perovskite grains buried within a complex, polycrystalline film is developed. The results of nXRD measurements are correlated with solar cell device measurements, and it is shown that spin-coating the perovskite precursor solution at elevated temperatures leads to improved surface coverage and enhanced solar cell performance.

## 1. Introduction

Since the seminal work of Kojima et al.[1] in 2009, in which organic-inorganic halide perovskites (OIHPs) were employed for the first time in dye-sensitized solar cells realizing a device with ~4% power conversion efficiency (PCE), research in the field of perovskite solar cells (PSC) has grown extensively. In 2012, Lee et al.[2] and Kim et al.[3] reported on mesoscopic perovskite solar cells with ~10% PCE. This triggered a research 'gold-rush' which has attracted considerable resources from both research groups and companies world-wide. Though first devices were based on mesoscopic structures in which the OIHP was included in a nanoporous titanium dioxide structure, it became clear that this structure was not required and that simple planar structures could provide very efficient solar cells.[4] Efforts recently culminated in the fabrication of devices with ~20% PCE in 2014[4] and ~21% in late 2015[5]; a performance that suggests OIHP could eventually replace standard crystalline silicon in photovoltaics. The broad range of optoelectronic properties, physical-chemistry aspects and low-production-cost potential of OIHPs have been widely reported in the literature.[4, 6-14] Here, we present an imaging method to simultaneously map the morphological and structural properties of the perovskites with a resolution of 400 nm, and focus on the importance of fine tuning the perovskite film morphology as a route to improve the PCE.

Three-dimensional OIHPs are described by the formula $ABX_3$, where A is a bulky monovalent cation (e.g. $CH_3NH_3^+$), B is a divalent metal halide cation (e.g. $Pb^{2+}$, $Sn^{2+}$, $Ge^{2+}$), and X is a monovalent halogen anion (e.g. $I^-$, $Br^-$, $Cl^-$). Structurally, the A cation is surrounded by a $BX_6$ octahedron, in which A is bonded to twelve X anions.[15, 16] To date, the most widely studied perovskite for PV applications has been methylammonium lead triiodide ($CH_3NH_3PbI_3$, which has been informally labelled MAPI). Several methods have been explored to fabricate perovskite layers, with solution deposition techniques under 'ambient' conditions being the most industrially attractive in terms of large area coating and low production cost.[8, 17-20] The simplest deposition method is the so-called 'one-step' process,[18, 21] in which a precursor ink is prepared by dissolving a mix of an ammonium salt (e.g. $CH_3NH_3I$, also known as MAI) and a lead salt (e.g. $PbI_2$, $PbCl_2$, or $PbAc_2$) in a solvent such as DMF. This ink can then be deposited using a variety of methods including spin-coating,[17] inkjet-printing,[19] and spray-coating.[18] The as-coated 'precursor' film is then thermally annealed to enable its conversion into the MAPI perovskite.[22] The perovskite layer is polycrystalline,[23, 24] with single crystallites having a typical lateral size ranging from a few tens of nanometers to 10-100s of microns, as evidenced by scanning electron microscope (SEM) imaging.[24-26] Perovskite grains are usually separated by voids or grain boundaries.[27] There has been some debate on the effect of grain boundaries,[28-30] with recent reports pointing towards the beneficial effect of their minimization[23, 24, 31, 32] as this reduces the defect density and thus charge traps, which leads to improved charge extraction and thus more efficient solar cells.[26] Achieving a uniform film coverage in which the polycrystals cover the entire solar cell active area without gaps or pin-holes[26] is another key morphological requirement to obtain efficient devices. Thus an ideal perovskite layer in a solar cell should consist of a smooth and flat single crystal without voids or grain boundaries. While this ideal morphology is hard to obtain by solution casting, larger grains and smoother films with fewer grain boundaries can be obtained by employing processing additives[33, 34] to the precursor solution[34,35], by using different solvent[35] and lead salts,[25] and by using solvent annealing.[23, 31]

To date, the most widely employed techniques to investigate the crystal structure and morphology of perovskite films have been SEM and X-ray diffraction (XRD). SEM has mainly been employed to image surface structure at both low and high magnifications. When combined with focused ion beam (FIB) it is also able to image cross-sections through complete solar cells, although this technique creates local damage. Perovskite crystallites can often be distinguished in as-spun precursor films; however, upon annealing for times longer than 20-30 min, these crystallites tend to merge in the final perovskite film.[6, 36] This makes the isolation of single crystallites and the determination of the statistics of lateral grain size and shape difficult.[36] A further limitation is that both SEM and other scanning probe microscopy techniques are surface-sensitive, with the bulk of the material being largely inaccessible using conventional imaging methods. The internal atomic scale structure of an OIHP is typically determined by X-ray diffraction techniques, among which synchrotron grazing incidence wide angle X-ray scattering (GI-WAXS)[37] is particularly popular. In the GI-WAXS technique, the X-ray beam is directed at an angle almost parallel to the sample surface. Large area detectors are employed to capture 2D diffraction patterns, yielding information about the structure both normal and parallel to the substrate. By varying the incident angle, depth-dependent structural information can be collected. In theory, this can be used to determine parameters such as the orientation of the crystallites, their lattice constants and average crystallite size. For this reason, GI-WAXS has proved to be a powerful

---

[1] Department of Physics and Astronomy, University of Sheffield, Hicks Building, Hounsfield Road, Sheffield S3 7RH, United Kingdom, e-mail: samuele_lilliu@hotmail.it, s.lilliu@sheffield.ac.uk
[2] European Synchrotron Radiation Facility, BP 220, Grenoble F-38043, France
[3] Cavendish Laboratory, University of Cambridge, Madingley Road, CB3 0HE Cambridge, United Kingdom
[4] Masdar Institute of Science and Technology, PO Box 54224, Abu Dhabi, United Arab Emirates
[5] School of Physics and Astronomy, Cardiff University, Queens Buildings, The Parade, CF24 3AA Cardiff, United Kingdom

technique for in-situ studies, where the effect of thermal annealing can be determined in real time.[36, 38, 39]

Conventional X-ray beams used in GI-WAXS have a cross-sectional area of 0.1 to 1 mm², and are thus orders of magnitude larger than the critical dimensions of most morphological features in OIHP films. This is exacerbated by the long footprint of the beam along the surface at grazing incident angles. For this reason, the observed diffraction patterns represent an ensemble average of structural features determined over a macroscopic area. Thus statistical information on the real-space distribution of crystallographic phases, lattice strain, crystallite orientation and disorder can mainly be inferred from peak shape analysis (which is extremely model-dependent), or is lost entirely. Fortunately, recent developments in X-ray focusing optics at high brilliance third generation synchrotron light sources now permit the routine use of intense, micron and submicron X-ray beams under ambient conditions.[40, 41] By raster scanning an X-ray 'nanobeam' across a sample, it is possible to combine real-space imaging with atomic scale structural information afforded by XRD. Such scanning nanofocus XRD (nXRD) techniques have been employed to resolve local variations in structure across a broad range of scientific fields including polymer and biopolymer fibers,[42] organic electronic materials,[43-45] macromolecular crystallography,[46] biological tissues,[47, 48] and semiconductor nanostructures.[49, 50] Scanning nXRD has also been extremely effective in the field of high-temperature superconductors.[51, 52] Campi et al. used nXRD to reveal the spatial distribution of charge-density-wave order and quenched disorder in $HgBa_2CuO_{4+y}$, a high temperature superconductor.[53] However, despite the wealth of structural information offered by nXRD, this technique has yet to be employed to examine the structure of OIHP films.

In this work, we demonstrate that nXRD can be used to gather spatially-resolved morphological and structural information on films of the archetypal $CH_3NH_3PbI_3$ perovskite in thin films that can be used directly in planar solar cell architectures. A method for selecting and visualizing grains diffracting according to a specific Miller plane has been developed using custom-made analysis software. Using this method, we show that nXRD is able to resolve the extent of individual perovskite grains buried within a polycrystalline film (grain segmentation). We then use nXRD, SEM and GI-WAXS to demonstrate that the perovskite film coverage across the substrate can be controlled by varying the temperature of both the precursor solution and the substrate during spin-coating. We conclude by showing that solar cells with the perovskite layer cast on a substrate held at a relatively high temperature result in devices with higher PCE.

## 2. Results and Discussion

We prepared three types of $CH_3NH_3PbI_3$ (MAPI) perovskite films by spin-coating a precursor solution made of methylammonium iodide ($CH_3NH_3I$ or MAI) mixed with lead chloride ($PbCl_2$) with a 3:1 molar ratio in N,N-Dimethylformamide (DMF), on $Si/SiO_2$ substrates coated with poly(3,4-ethylenedioxythiophene) polystyrene sulfonate (PEDOT:PSS). The cold-spun samples (here termed 'cold') were spin-coated on a substrate held at room temperature. The medium-spun samples ('medium') were spin-coated with the substrate at ~75°C. The hot-spun samples ('hot') were spin-coated with the substrate at ~90°C. In all cases, the temperature of the precursor ink was held at 75°C before it was spin-cast, with all coating performed under ambient conditions (lab temperature of ~24°C at a relative humidity of ~30%). The samples were then annealed under ambient conditions for 90 min on a hotplate set to a temperature of 90°C, and kept in nitrogen before the measurements. The film thicknesses for the cold, medium, and hot samples measured with a profilometer after annealing were ~350 nm, ~500 nm, and ~650 nm, respectively. Further details are available in the Experimental Section.

### 2.1 Scanning Electron Microscopy (SEM)

Typical SEM images of annealed cold, medium and hot MAPI films are shown in **Figure 1**. It can be seen that improved film coverage is obtained at increased substrate temperatures during spin-coating. This result confirms previous reports that have shown that casting the precursor onto a relatively warm substrate can assist the formation of a more uniform morphology.[18, 54] Here, the creation of a uniform precursor film is important for producing a uniform perovskite film; a result likely explained by the reduction in film volume caused by thermal annealing.[36] As seen in Figure 1, the perovskite grains appear to have melted together,[36] making it difficult to define the location of the grain boundaries; an effect that limits grain segmentation analysis. It is also clear that the films are multilayered, although it is not possible to resolve such sub-surface structure using SEM.

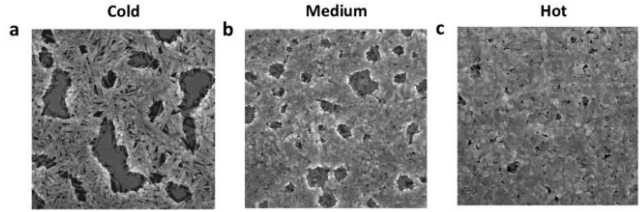

**Figure 1.** SEM scans of cold, medium and hot spun MAPI films. In each case, the scan area is 40 × 40 µm². It can be seen that the film surface coverage increases as the deposition temperature is increased.

### 2.2 Grazing Incidence Wide Angle X-Ray Scattering (GI-WAXS)

Cold, medium and hot spun MAPI films were probed by grazing incidence wide angle scattering (GI-WAXS) at the BM28 beamline (ESRF, Grenoble, France) (see Experimental Section for details). **Figure 2** shows GI-WAXS diffraction patterns recorded for three samples measured above the critical angle ($\alpha_c \approx 0.16°$) at an out-of-plane incident angle equal to $\alpha_i \approx 0.3°$, achieving a penetration depth of ~140 nm (see Figure S1 Supporting Information). Close inspection of the Debye-Scherrer rings[55] evident in Figure 2 indicates that they are in fact composed by thousands of diffraction spots,[36] as can be seen in the zoom in the insets. We plot further in-plane and out-of-plane line profiles at different incident angles in Figures S2 and S3. Interestingly, these measurements do not reveal any significant structural differences between the three samples nor any information on depth-related structural variation. It is likely, however, that lattice strain and sample size effects or, equivalently, beam footprint effects (see Figure S4) play a significant role in peak broadening. All these effects add up, making it difficult to quantify the lattice constants and the domain size of this class of polycrystalline materials using GI-WAXS. Note, however, that GI-WAXS is an extremely powerful technique when tracking relative changes (orientation and peak position) of individual diffraction spots in-situ during a thermal annealing.[36]

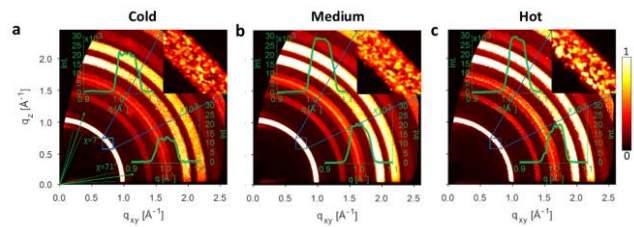

**Figure 2.** GI-WAXS diffraction patterns of cold, medium and hot spun MAPI films. The arrows in part **a** indicate the azimuthal angle ($\chi$) of the cake slice used for extracting the out-of-plane line profile ($\chi = 7°$) and the in-plane line profile ($\chi = 71°$), which are both shown in each figure. The dotted lines indicate the cake slice aperture ($\chi = 10°$) that is used in the line profile integration. The images in the top corner represent a zoom of the square region of interest indicated in blue. Diffraction patterns are normalized between 0 and 1 and are represented on a linear scale.

### 2.3 Scanning Nanofocus X-Ray Diffraction (nXRD)

As we show below, the limitations of both SEM and GI-WAXS outlined above can be overcome using Scanning Nanofocus XRD (nXRD), which we use to study the same perovskite films deposited at different temperatures. Measurements were performed at the ID13 beamline (ESRF, Grenoble, France), where each sample was placed on a holder mounted on high-speed *xyz* piezo scanning stages. An optical microscope with a 50× magnification was first used to image a specific region of interest and thus define the centre of the nXRD scan (**Figure 3**a). Figure 3c, f, i show the optical micrographs corresponding to the region in which nXRD measurements were performed.

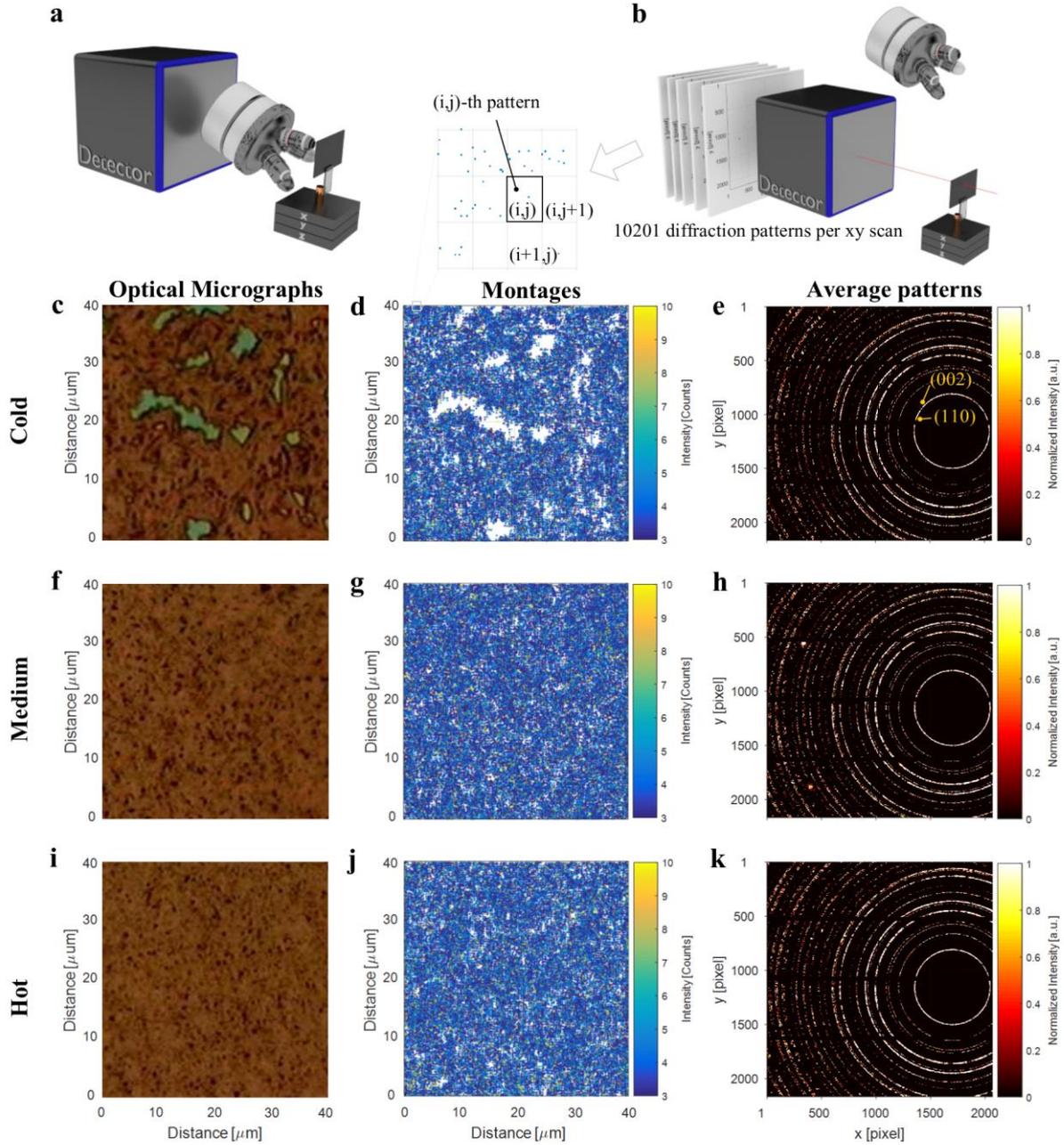

**Figure 3.** Summary of measurements performed with Scanning Nanofocus XRD (nXRD). a, nXRD scan area selection with the optical microscope. b, Setup during nXRD measurements. The inset in b illustrates the construction of a montage, where '(i,j)-pattern' indicates single diffraction patterns from the 10201 diffraction pattern dataset in a raster scan. c, f, i, optical micrographs acquired before the scan on the scan area for the cold, medium, and hot samples. d, g, j, nXRD montages for the cold, medium, and hot sample. e, h, k, average diffraction patterns recorded from the cold, medium, and hot sample.

The microscope was then retracted (Figure 3b) and the sample was illuminated in transmission mode with a monochromatic beam (Energy ≈ 14.85 keV) focused into a $200 \times 200$ nm$^2$ spot size, which was raster scanned over a $40 \times 40$ µm$^2$ area with a step of ~400 nm along the $x$ and $y$ directions. Diffraction patterns were collected at each point of the raster scan using a 2D detector. A full scan consisted of 10201 diffraction images. Further details are available in the Experimental Section. Pictures of the setup are shown in Figure S5 Supporting Information.

To analyse the data, we developed a MATLAB application (see Experimental Section and Figure S6). The 10201 diffraction patterns of a raster scan were de-noised as described in the Supplementary Information and Figure S7 and S8. Montages of the diffraction images were constructed by placing each diffraction image at the spatial coordinate at which it was collected (see inset Figure b and Figure S9). The sparse patterns of these montages are shown in Figure 3d, g, j for the three samples.

Note that the development of software that could handle sparse matrices was essential for this step. In fact, a montage of $101 \times 101$ images (each of 17 MB in the standard synchrotron *.EDF format), would have resulted in a ~170 GB image, which could not have been easily visualized. Instead, by displaying the montage as a sparse pattern we can explore multiple diffraction patterns in real time from a chosen region of interest. The construction of a montage allows us to immediately establish a correlation between the structural texture of the film and its morphological properties (approximating the local intensity of X-ray scattering as is evident in Figure 3). This is particularly evident for the cold sample. Here, the optical micrograph (Figure 3c) shows clear discontinuities in the film, with the green regions corresponding to the underlying Si/SiO$_2$ substrate. The same regions can be seen in the corresponding montage (Figure 3d) and correspond to zones in which no diffraction data were collected (plotted using white colour on the montage image). The surface coverage, quantified as the ratio between the images that contain diffraction data and the total number of images in a scan (10201), is ~92% for the cold sample. In contrast, the medium and hot samples are characterized by a much higher degree of surface coverage (~99%). Interestingly, in these films, the montages reveal details that are not evident from the optical micro-graphs. In fact, from the optical micrographs, one would conclude that there is an almost perfect coverage in the medium and hot samples; however, the montages show small voids, corresponding to regions in which no diffraction was recorded.

We can also construct an average diffraction pattern from the diffraction patterns recorded during the raster scans as shown in Figure 3e, h, k for the three samples. This average diffraction pattern is similar to those usually collected in a single scan by using a beam having a footprint 200 times larger (still one order of magnitude smaller than in a regular GI-WAXS measurement). When recording diffraction patterns in GI-WAXS using much larger beams, a larger number of grains are illuminated, with such grains having different relative separations from the detector. Here, single grains are probed in transmission with a nanofocussed beam at a fixed grain-detector separation. As a consequence, the Debye-Scherrer rings in the nXRD average pattern are not affected by sample-size broadening and are significantly sharper than the Debye-Scherrer rings measured in GI-WAXS geometry.

It can be seen in Figure 3e, h and k that the average scattering patterns for the cold, medium and hot samples are very similar. To analyse this in more detail, we extract azimuthally integrated line profiles from the average patterns. The process by which this is done is described in the Supporting Information and is illustrated in Table S1 and Figure S10, with the azimuthally integrated line profiles shown in Figure S11. As in other reports,[56, 57] we can distinguish between the (110) and (002) peaks, which are usually classified as the first main perovskite peak. Indeed, in other GI-WAXS reports, this two double peak is often simply identified as the (110) peak.[6, 25, 58-60] This is likely due to poor instrumental resolution that does not allow these two reflections to be resolved. One should note that such double peaks converge into single peaks ((100) and (200)) at temperatures greater than 54-57°C, when the perovskite phase converts from a tetragonal ($\beta$-phase) to a cubic ($\alpha$-phase).[56, 61] Clearly, approximating such double peaks as single peaks inevitably leads to inaccuracies in the calculation of crucial parameters such as lattice constants.

Since both MAI:PbCl$_2$ and MAI:PbI$_2$ precursor solutions crystallize as CH$_3$NH$_3$PbI$_3$ (MAPI),[25] the line profiles shown in Figure S11 were used to refine the MAPI tetragonal structure from ref.[62] (Figure S12 shows a comparison between measured and simulated patterns, and Tables S2-S4 show the indexed peaks). On the basis of our measurements, we determine a = b ≈ 8.88 Å and c ≈ 12.66 Å as the average lattice parameters for the three samples (Table S5), which is in good agreement with the refined $\beta$-phase[61] MAPI crystals reported by Liang et al. (a = b ≈ 8.874 Å and c ≈ 12.670 Å),[57] Im et al. (a = b ≈ 8.883 Å and c ≈ 12.677 Å),[63] Kojima et al. (a = b ≈ 8.855 Å and c ≈ 12.659 Å).[1] Other authors have reported a shorter c lattice constant for perovskites obtained from the PbCl$_2$ lead salt (c ≈ 11.24 Å),[2, 6, 64] which was explained as resulting from the incorporation of Cl in the lattice structure.[64] Others reported a minor difference between perovskites obtained from PbI$_2$ and PbCl$_2$ lead salts (c ≈ 12.67 Å for PbI$_2$ and c ≈ 12.64 Å for PbCl$_2$), which was attributed to a degree of Cl-doping in the film.[65] Discrepancies between these and our measurements could be due to different instrumentation and processing methods (e.g. annealing temperature and time), different MAI batches, or sample treatment (note that in ref.[2] the diffraction patterns show diffraction peaks from both Pb and from Cl). From our measurements we determine that perovskites deposited from PbI$_2$ and PbCl$_2$ lead salts have the same MAPI crystal structure.[25] Due to the similarities between the average structural data of the three samples, we therefore conclude that the different processing temperatures explored here do not affect the unit-cell crystal structure of the perovskite films.

We now focus on the texture of the three samples and illustrate a method for performing grain segmentation and quantitative analysis. Here we focused on the (002) and (110) reflections, although the method can be extended to higher order reflections. To do this, a circular region of interest was defined that included only these reflections (Figure S13). (Note that in the following discussion the terms 'image' and 'diffraction image' refer to the circular region of interest.) Diffraction spots in each image were identified and assigned to (002) or (110), and then analysed. For each spot peak a double Gaussian line profile was drawn through its line profile, from which values of ($q_p$, $\chi_p$, and intensity $I_p$) were extracted (see Figure S14). The spatial coordinates of the $i$-th scan and $\chi_p$ of the $j$-th diffraction spot were then used to perform grain clustering in an attempt to identify distinct grains having a specific $\chi$ orientation (see Supporting Information). The assumption made in the clustering procedure is that diffraction spots that are adjacent both in spatial coordinates and in reciprocal space coordinates most likely originate from the same grain.

Once single grains were clustered, spatial maps of $q_p$, $\chi_p$, intensity, FWHM, etc. were constructed (see Figure S15-S21), allowing us to determine the grain size with an accuracy of ±200 nm based on the spatial location of the diffraction spots. The grain size reported here corresponds to the lateral area (parallel to the substrate and perpendicular to the X-ray beam) of the perovskite platelets. From the clustered diffraction-spot maps, we also constructed quiver-plots. In a quiver plot the value of $\chi_p$ determined is represented using an arrow with its centre located in the spatial position from which the diffraction spot was acquired, and with an orientation and colour corresponding to $\chi_p$. Such quiver plots usefully allow multiple arrows to be visualized, even if they are centred at the same location; a situation that occurs due to the overlap of grains having different values of $\chi_p$. Using the clustering and visualization tools developed here, we are able to selectively classify grains having a specific sizes or other properties. In **Figure 4** we show quiver plots generated for the (002) and the (110) reflections for the cold, medium and hot films, where the thicker arrows, which at the present zoom appear as dots (magnifications are available in Figure S22-S33).

Although all data are plotted in this figure, quiver arrows are highlighted for grains having a size larger than 4 μm$^2$ using a colour scale that indicates the relative orientation of to $\chi_p$. We also identify largest grain imaged in each sample using a white rectangular box. It can be seen that the number of grains having a lateral size larger than 4 μm$^2$ apparently correlates with increasing spin-coating temperature. We present statistics on grain size in Figure S34 and Table S6, which confirm that the maximum grain size increases as the spin-coating temperature increases. Extracting the distribution of grain sizes, exponential in this case, provides more information than the usually quoted average grain size determined from the measured peak width. It also gives an average grain size that is not affected by microstrain, lattice distortions, inhomogeneity and instrumental broadening contributions that can affect peak width analysis.

From Figure 4d, g, j, m and p it can be seen that the largest grains in each quiver plot often overlap with other grains. The degree of overlap, i.e. the number of different grains that exist at the same spatial position, can be quantified by calculating the average number of diffraction spots per diffraction image (i.e. circular region of interest). We find that this number to be low in the cold cast sample (1.44 ± 0.71 for (002) and 1.42 ± 0.70 for (110)), and higher in the medium (1.46 ± 0.72 for (002) and 1.46 ± 0.72 for (110)) and hot sample (1.57 ± 0.85 for (002) and 1.57 ± 0.83 for (110)). This value, in fact correlates well with film thickness; i.e. the thicker samples are characterized by more overlapping crystallites. Although we are unable to determine the thickness of these grains, the ability to visualize overlapping (or buried) grains and grains with specific properties, such as size or orientation, makes this technique extremely powerful.

From the (002) $q_p$ values, we extracted the c lattice constant using c = $4\pi/q_p$. Similarly, we extracted the a = b lattice constants from the (110) $q_p$ values using a = b = $2^{0.5}\pi/q_p$ (tetragonal crystal). We show spatial maps of the lattice constants for the largest (002) and (110) grains (highlighted by using the white rectangles) for the cold, medium and hot samples in Figure S19-S21 c and d, respectively. These maps are not flat and apparently indicate a non-uniform distribution of lattice constants within an individual grain. We can quantify the percentage deviation of the average lattice constant within the same grain using $\epsilon_c = 100 \times (q_{p(002)} - q_{p,av(002)})/ q_{p,av(002)}$ and $\epsilon_a = 100 \times (q_{p(110)} - q_{p,av(110)})/ q_{p,av(110)}$, where $q_{p(002)}$ is the (002) $q_p$, $q_{p(110)}$ is the (110) $q_p$, $q_{p,av(002)}$ is the average (002) $q_p$ within the grain, and $q_{p,av(110)}$ is the average (110) $q_p$. Spatial maps of $\epsilon_c$ and $\epsilon_a$ are shown in Figure 4, indicating the presence of strain across single crystals. This is further illustrated in Figure 4, where we plot normalized Gaussian fits corresponding to the (002) and (110) reflections for the largest grains (where each Gaussian fit corresponds to a single pixel from the selected grain). Note that these fits are normalized to make all curves visible (raw data fits are shown in Figure S35).

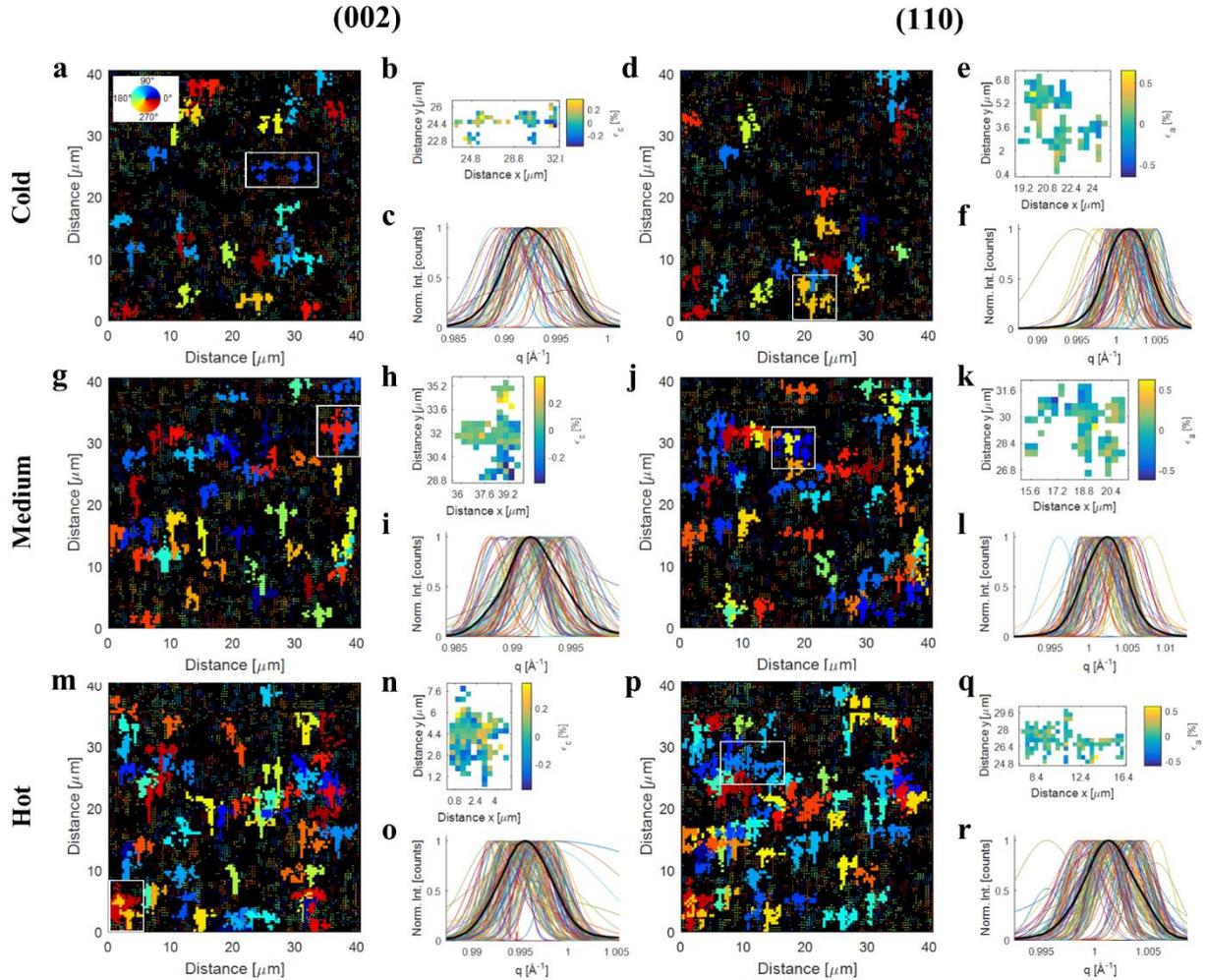

**Figure 4.** Quiver plots highlighting (002) and (110) perovskite grains, with $\epsilon_a$ and $\epsilon_c$ maps presented for grains larger than 4 μm².

In Figure 4 we also plot the normalized sum profile (displayed as thick black line) that would have been approximately obtained if the X-ray beam had been defocused to illuminate an entire grain. As an example, from analysis of the average sum profile in Figure 4c we find an FWHM of 0.0066 Å⁻¹ and an average FWHM of the single line profiles of 0.0050±0013 Å⁻¹. This is equivalent to ~24% strain contribution to peak broadening (similar calculations for the largest grains are available in the Table S7). Unfortunately, we are unable to correlate the (002) or (110) reflections identified with higher order reflections to perform a Williamson-Hall plot[66-68] within the same grain. This could theoretically be achieved by rotating the sample about an ideal rotation axis coincident to the X-ray beam. Using this technique, it is possible to establish which diffraction spots belong to the same crystal (see the 3D-XRD technique for example[69, 70]). However, combining this with an nXRD setup is technically challenging, as it requires the use of an axial rotation stage together with nXRD scanning stages.[36] Furthermore, such measurements would be compromised by the degradation of the perovskite film, as individual crystallites are unlikely to 'survive' for more than 5 s when continuously exposed at the same spot.[36]

### 2.4 The Role of Substrate Temperature on Solar Cell Efficiency

Finally, we investigated the effect of substrate deposition temperature on the photovoltaic performance of a perovskite solar cell. Solar cells were fabricated and measured as described in the Experimental Section. The device architecture used was glass/ITO/PEDOT:PSS (30nm)/perovskite/PCBM (120nm)/Ca (5nm)/ Al (100nm). Solar cells were measured under standard light conditions (100 mW/cm², AM1.5G illumination), without initial light soaking and without pre-biasing. In **Figure 5** we plot current-density vs. voltage curves recorded from the devices that were deposited from a precursor solution on to cold, medium or hot substrates. To ensure statistical significance, we measured between 11 to 16 pixels for each deposition condition (note, one solar cell only contained 4 useful pixels). **Table 1** displays average and champion device metrics for the different conditions, recorded using forward (-1 to +1) and reverse (+1 to -1) JV scans. It can be clearly seen that the solar cell performance is maximized when devices are deposited onto a hot substrate (a champion (average±std) PCE of 12.8% (11.9±0.8)% is obtained from the reverse scan). We find that the improvement in device efficiency mainly occurs as a result of improved $J_{sc}$, a result consistent with improved film coverage and enhanced optical absorption occurring in precursor films deposited onto a hot substrate.

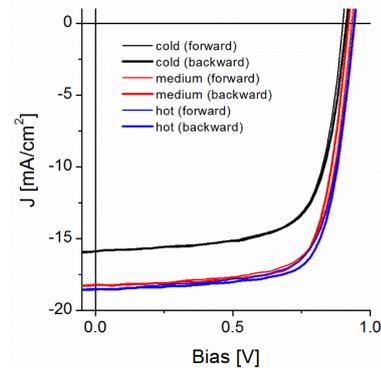

**Figure 5.** Current-density vs. voltage measurements for the best cold, medium and hot spun perovskite solar cells. The measurements were recorded without any light soaking, both from forward (from -1 to 1V) and reverse (from 1 to -1V) voltage sweeps.

**Table 1.** Spin-coated solar cells performance. Average, standard deviation, and maximum values (in brackets) of the solar cell figures of merit measured on at least 11 solar cells for each cell type.

|  |  | Number of pixels | PCE [%] | $J_{SC}$ [mA/cm2] | $V_{OC}$ [V] | FF [%] |
|---|---|---|---|---|---|---|
| Cold | Forward | 16 | 9.5±0.3 (9.9) | 15.4±0.4 (16.1) | 0.882±0.013 (0.902) | 70.2±1.3 (71.9) |
| Cold | Backward | 16 | 9.6±0.3 (10.2) | 15.4±0.4 (16.1) | 0.890±0.014 (0.913) | 70.0±1.7 (71.9) |
| Medium | Forward | 12 | 11.2±0.8 (12.2) | 17.7±0.3 (18.2) | 0.900±0.023 (0.932) | 70.0±2.2 (73.1) |
| Medium | Backward | 12 | 10.8±0.7 (12.3) | 17.5±0.5 (18.2) | 0.895±0.011 (0.920) | 68.9±2.6 (73.4) |
| Hot | Forward | 12 | 11.5±0.8 (12.3) | 18.2±0.3 (18.6) | 0.910±0.030 (0.951) | 69.3±2.6 (72.9) |
| Hot | Backward | 11 | 11.9±0.8 (12.8) | 18.3±0.3 (18.7) | 0.917±0.028 (0.953) | 71.1±2.2 (74.0) |

## 3. Conclusion

We have employed synchrotron Scanning Nanofocus X-ray diffraction (nXRD) to probe the morphology and the structural properties of spin-coated $CH_3NH_3PbI_3$ (MAPI) films used in solar cell applications. In particular, using nXRD we are able to unambiguously perform grain segmentation and extract the lateral size, strain and orientation of individual perovskite platelets. In contrast to SEM, nXRD allows the full depth of the film to be spatially imaged, identifying overlapping grains at different depths. Furthermore, nXRD allows us to identify the distribution of grain sizes and their orientation, together with the spatial variation of strain along the a and c directions within individual grains. In contrast GI-WAXS diffraction images for the cold, medium and hot samples show very few discernible differences between films. Our nXRD measurements also show that the substrate temperature during perovskite precursor film spin-coating affects the final coverage of the annealed perovskite film. We found that substrates spun at relatively high temperature, result in a better film coverage and larger grain sizes, which in turn created solar cell devices having improved light harvesting abilities and thus enhanced power conversion efficiency.

## 4. Experimental Section

Some of the details regarding sample and solar cell preparation as well as GI-WAXS setup have already been reported in our previous works[18, 36, 71] and are reported here again for clarity.

### 4.1 SEM, GI-WAXS, and nXRD Samples Preparation

$Si/SiO_2$ substrates (Ossila Ltd (UK), code S143) were cleaned by sonication in isopropyl alcohol and deionized water (10 min each), and dried with compressed nitrogen before use. Methylammonium iodide (MAI) powder was synthetized as per our previous work.[18] Lead II chloride ($PbCl_2$, 98% purity) was purchased from Sigma Aldrich (268690, Aldrich). Precursor solutions of MAI and $PbCl_2$ (3:1 molar ratio) were dissolved in sequence into dimethylformamide (DMF) with a concentration of 664 mg ml$^{-1}$, heated at ~75°C overnight to facilitate dissolution of solid material, cooled to room temperature, and then filtered through a 0.45 mm PTFE filter before use. Samples were prepared in ambient conditions (lab temperature ~24°C, relative humidity ~30%). A ~40 nm layer of poly(3,4-ethylenedioxythiophene) polystyrene sulfonate (PEDOT: PSS, Heraeus Clevios™ P VP AI 4083 from Ossila, code M121) was spin-coated (5000 rpm, 30 s) on the $Si/SiO_2$ substrate immediately before the deposition of the perovskite layer. Three sets of samples were prepared: cold, medium, and hot spun samples. The 'cold spun' samples were spin-coated on a substrate held at room temperature. The 'medium spun' samples were prepared by spin-coating a substrate transferred from a hotplate at ~90°C. The time required for transferring the substrate from the hotplate to the spin coater chuck was 5 s. Using a thermometer with a k-type wire thermocouple, we estimated that the substrate at the beginning of the spin-coating was at ~75°C. The 'hot spun' sample was prepared as the 'medium spun' sample, but with the hotplate set at ~120°C. In this case, the substrate at the beginning of the spin-coating was at ~90°C. The as-prepared samples were sealed inside a glovebox a shipped via air fright under controlled conditions for the GI-WAXS and nXRD measurements.

### 4.2 SEM

SEM images were taken with an FEI Nova NanoSEM. We used a 20-30 kV incident beam and a secondary electron detector.

### 4.3 GI-WAXS

Details on the BM28 (XmaS, ESRF, Grenoble, France) beamline are available at www.xmas.ac.uk. The radiation coming from a bending magnet (critical energy $E_c$ = 9.8 keV) was monochromatised using a fixed-exit, water-cooled, double crystal Si(111) monochromator, placed at 25 m from the source. A Rh-coated toroidal mirror was used to focus the monochromatic beam horizontally and vertically. The X-ray beam was focused to a beam spot size of ~500 × 100 µm², resulting in beam footprint of ~500 × (100 sin($\alpha_i$)) µm², where $\alpha_i$ is the (out-of-plane) incident angle. Cold, medium, and hot spun films were prepared as described above and spin-coated at the beamline, annealed, and measured within one day. Here, samples were housed in a helium-filled custom made environmental chamber mounted on an 11-axis Huber diffractometer.[36] A MAR 165 CDD detector (2048 rows × 2048 columns) was placed on the diffractometer arm at a distance of 223 mm from the sample (calibrated as in ref. [72, 73]). Single diffraction patterns were obtained by illuminating the samples for 10 s with an X-ray beam having a wavelength of ~1.2384 Å. Data processing was performed using Matlab software.[74]

For the calculation of the GI-WAXS penetration depth we used the following expression:[37]

$$\Lambda \approx \left[ \sqrt{2}k \sqrt{\sqrt{(\alpha_i^2 - \alpha_c^2)^2 + 4\beta^2} - (\alpha_i^2 - \alpha_c^2)} \right]^{-1}$$

where k = 2π/λ is the wavenumber (λ is the X-ray wavelength), $\alpha_c$ is the critical angle, $\alpha_i$ is the out-of-plane incident angle, and β is the imaginary part of the index of refraction. The index of refraction is defined as n = 1 − δ + iβ.[37] The critical angle was calculated as $\alpha_c$ = √(2δ). The δ and β parameters for $CH_3NH_3PbI_3$ at λ = 1.2398 Å were estimated from the online toolbox available at ref. [75], and are equal to 3.76813 × 10$^{-6}$ and 3.06721 × 10$^{-7}$, respectively. A plot of the penetration depth for these experiments is shown in Figure S1, Supporting Information.

### 4.4 nXRD Beamline Setup

Nanofocus XRD (nXRD) measurements were performed at beamline ID13 (ESRF, Grenoble, France). A monochromatic beam of λ = 0.8349 Å (Energy $E$ = 14.85 keV) was focused by crossed linear refractive silicon nanofocusing lenses (NFLs)[76] to a spot size of ~200 × 200 nm², with an incoming flux of approximately 5 × 10$^{10}$ photons/s. Additional lead shielding and an electron microscopy aperture (20 µm diameter) were used to remove any parasitic background scattering. All measurements were performed under ambient conditions (lab temperature ~24°C, relative humidity ~40%).

Samples were mounted on a high speed *xyz* piezo scanning stages on top of a coarse positioning 6-axis hexapod (Physik Instrumente) such that the perovskite material was facing downstream of the incoming X-ray beam. Samples were positioned and aligned to the focal plane of the X-ray beam with an on-axis microscope (see Figure S5 Supporting Information). This microscope was also used for selecting regions of interest for measurements and recording optical micrographs. Data were recorded on an EIGER 4M detector (Dectris) with 2168 rows × 2070 columns and pixel size of 75 × 75 µm², using an exposure time of 0.1 s. The detector was placed at a distance of 0.1941 m from the sample. The detector position and geometry were calibrated by recording a diffraction pattern of the standard calibration material corundum ($\alpha$-$Al_2O_3$) and using the *pyFAI-calib* calibration routine, which yielded the distance, point-of-normal-incidence and detector rotation angles.[77]

### 4.5 nXRD Measurements and Data Analysis

During data collection, samples were scanned within the *y-z* laboratory plane (with the X-ray beam along the *x*-axis), here indicated as *x-y*. A picture of the beamline configuration used during the nanofocus raster scan is shown in Figure S5 Supporting Information. Each scan consisted of a 40 × 40 µm² raster scan of 101 × 101 pixels, with a step size of 400 nm. A single scan resulted in 10201 2167 × 2070 pixel diffraction images, for a total of ~170 GB of data saved in the ESRF Data Format (EDF). Data Analysis was performed with a MATLAB[78] application specifically designed by S. Lilliu for nXRD measurements (Figure S6). We used a custom NNB-XB0057 (NOVATECH, UK) notebook with a quad-core Intel® Core™ i7-4790K CPU 4GHz, 32GB of RAM, and Windows 7 (64bit).

Single diffraction patterns were affected by noise in the form of 1-2 pixel counts randomly distributed in the image (Figure S7a). Once these pixels were set to zero, single patterns consisted of highly sparse matrices[79] (Figure S7b). As an example, the average sparsity or density[79] of the 10201 diffraction images of the cold scan was $(1.9 \pm 1.8) \times 10^{-3}$%. We found that the fastest way to read a single diffraction pattern was to create a memory-map[78] of each EDF file and then convert the mapped file into a double matrix variable. This matrix was then converted into a sparse matrix format,[78] in which data were represented as list of three elements: (row, column, counts or, equivalently, intensity). Representing matrices in the sparse format saved significant amounts of storage space and resulted in faster data processing. The time taken for data conversion was mainly limited by the low data speed transfer between the external hard drive and the processor (~50 MBps). Therefore, parallel processing did not result in any improvement. The conversion process, where the 10201 image were saved in a struct[78] variable, took approximately 1 hour. This step compressed data ~$2.4 \times 10^4$ times (from ~170 GB to ~7 MB).

The average diffraction pattern of the 10201 images was calculated (in ~10 s) as the sum of the diffraction patterns divided by the number of diffraction images in a scan. From the average diffraction pattern we observed some residual noise and broad diffused scattering from the substrate ($Si/SiO_2$). We therefore reprocessed all the diffraction patterns by applying a 'clean' filter, which set to zero all the pixels surrounded by zeroes (Figure S7c). A comparison between the average diffraction pattern before and after the application of the filter is shown in Figure S8.

Scan 'montages' were constructed by concatenating the 10201 patterns of a single scan in a 101 × 101 frame consisting of a large 218867 × 209070 sparse matrix. Spatial coordinates were assigned to the centre of each diffraction pattern according to the raster scan. The concept is illustrated in Figure S9 for 25 spatially adjacent diffraction patterns (cold sample). With a 40 × 40 µm² scan area, the 101 × 101 scan points split the scan area into 0.4 × 0.4 µm² tiles. To avoid negative values, we centred the first scan point (1,1) or tile at (0.2, 0.2) µm. Therefore, the last point of the scan is located at (40.2, 40.2) µm.

A diffracting sample projects a cone of rays (Debye-Scherrer rings[55]) onto the detector. The projection of this cone onto a flat detector perpendicular to the direct beam is a circle. However, for various reasons, the detector might not be perfectly perpendicular to the direct beam. As a result, the rings appear on the acquired image as tilted ellipses, rather than circles.[73] The 'calibration' step consisted in extracting the parameters required for correcting this distortion and converting the row and column image coordinates into scattering vector coordinates.[73] The geometry employed in the calibration is described in ref.[80]. Measured diffraction patterns from the α-$Al_2O_3$ were measured before the experiments. With the *pyFAI-calib* library[77, 81, 82] we extracted the distance between the sample and the detector, 3 rotation angles (rot1, rot2, rot3), and the points of normal incidence (PONI),[80] which were used in the conversion of the diffraction image coordinates from rows and columns to scatting vector polar coordinates, as discussed in the next section. The extracted parameters are reported in Table S1.

These parameters were used for converting row and column image coordinates into scattering vector coordinates in the following way ('remapping' step). We first calculated the detector rotation matrix **R**

$$\mathbf{R}_1 = \begin{bmatrix} 1 & 0 & 0 \\ 0 & \cos\rho_1 & -\sin\rho_1 \\ 0 & \sin\rho_1 & \cos\rho_1 \end{bmatrix} \quad (1)$$

$$\mathbf{R}_2 = \begin{bmatrix} \cos\rho_2 & 0 & \sin\rho_2 \\ 0 & 1 & 0 \\ -\sin\rho_2 & 0 & \cos\rho_2 \end{bmatrix} \quad (2)$$

$$\mathbf{R} = \mathbf{R}_2 \mathbf{R}_1 \quad (3)$$

where $\rho_1$ = -rot1, $\rho_2$ = -rot2, rot1 and rot2 are the detector rotation angles extracted from the calibration (see **Error! Reference source not found.**). We then calculated:

$$p = \begin{bmatrix} c - \text{PONI}c \\ r - \text{PONI}r \\ L \end{bmatrix} \quad (4)$$

where $c$ and $r$ are the column and row coordinates of a non-zero pixel in the diffraction pattern, PONI$c$ is the column coordinate of the point of normal incidence,[81] and PONI$r$ its row coordinate. The point $p$ was rotated through the detector rotation matrix:

$$t = \begin{bmatrix} t_1 \\ t_2 \\ t_3 \end{bmatrix} = \mathbf{R}p \quad (5)$$

The Bragg angle in radians $2\theta$ is:[55, 80]

$$2\theta = \text{atan2}\left(\sqrt{t_1^2 + t_2^2}, t_3\right) \quad (6)$$

The norm of the scattering vector $q$ is:[55]

$$q = \frac{4\pi}{\lambda} \sin\frac{2\theta}{2} \quad (7)$$

And the azimuth in radians is:

$$\chi' = -\text{atan2}(t_1, t_2)$$
$$\chi = \begin{cases} \chi' & \text{for } \chi' > 0 \\ \chi' + 2\pi & \text{for } \chi' < 0 \end{cases} \quad (8)$$

For easier data processing and visualization we used $\chi$ (instead of $\chi'$) in degrees (from 0° to 360°). Therefore, the 'remapping' step remapped a non-zero (row, column, intensity) triplet into a ($q$, $\chi$, intensity) triplet. In Figure S10 we show the $q$, $\chi$ coordinates, or simply polar coordinates (distance and azimuth), for a diffraction spot.

The extraction of azimuthally integrated line profiles required two further steps: rebinning and averaging. The intensities of the ($q$, $\chi$, intensity) triplet were average-binned into a sparse matrix having a regular step size for $q$ and $\chi$ ($\Delta q = 1 \times 10^{-3}$ Å$^{-1}$ and $\Delta\chi = 0.5°$). The choice of the step size had to take into account the $q$ resolution. In fact, the size of pixels close to the direct beam ($q \approx 0$ Å$^{-1}$) is $\Delta q \approx 3 \times 10^{-3}$ Å$^{-1}$, while the size of pixels close to the left edge of the diffraction pattern ($q \approx 4.3$ Å$^{-1}$) is $\Delta q \approx 2 \times 10^{-3}$ Å$^{-1}$. Therefore, we can set the resolution to $\Delta q \approx (2.5\pm0.5) \times 10^{-3}$ Å$^{-1}$. As a consequence, a step size of $\Delta q < 1 \times 10^{-3}$ Å$^{-1}$ would result in oversampling and noisy data.

In the binning step we collected all the elements from the triplet that fell into the (m,n)-th bin, i.e with $q$ coordinates between m$\Delta q$ and (m-1)$\Delta q$ and $\chi$ coordinates between n$\Delta\chi$ and (n-1)$\Delta\chi$, and calculated the average of the intensity of these elements. The obtained 'polar' matrix $I$(q,$\chi$) was then used for the calculation of the 'average azimuthally integrated line profiles':

$$i(q) = \sum_{\chi=0°}^{360°} \frac{I(q,\chi)}{N_q} \quad (9)$$

where, here, the azimuthal integration was performed from 0° to 360°, and $N_q$ is the number of elements with the same $q$. In practice, the line profile was simply calculated as mean(I) in MATLAB. Line profiles extracted for the cold, medium, and hot sample are shown in Figure S11.

With our MATLAB application it is possible to inspect single diffraction patterns in real time by simply selecting the image index, the column and row of the raster scan, or by simply pressing up/down or left/right arrows. The line profile extraction, which includes 'remapping' and 'binning' is performed in real time. The processing time for these two steps is ~1.2 ms per image (average value calculated on the 10201 images of the cold sample). Note that the processing speed is inversely proportional to the density of the sparse matrices.

Parallel processing or GPU processing[78] was not an option for highly sparse dataset, as the time taken to dispatch/collect the job or to generate a GPU variable and retrieve results after processing was longer than the actual processing time.

The line profile extracted from the average diffraction pattern of the three samples were converted into 2θ coordinates. The MAPI crystallographic file used for the Rietveld refinement[83] was obtained from ref.[62] The pattern was fitted with Pseudo-Voigt functions.[37]

To process data within a certain $q$ range we defined circular regions of interest (ROIs). Figure S13 shows an example of ROIs including data from the (002) and the (110) reflections, from the average pattern. Here we used the average pattern just to display the concept of circular ROI, as single diffraction patterns would only display a few scattered diffraction spots. We first generated a binary mask by defining an annulus in a 2167 × 2070 (which is the same size of a single diffraction image) null matrix with a large radius R = 355 pixels and a small radius r = 336 pixels and assigning 1 to the pixels inside it. The binary mask was converted into a sparse matrix and multiplied point by point to the (i,j)-th diffraction pattern of the raster scan. This allowed us to consider only data from the (002) and (110) reflections.

The analysis followed two conceptual steps: (i) diffraction spots clustering in the single diffraction patterns and extraction of their features including the peak position, (ii) spots clustering on the full dataset based on the spatial distance in the raster scan and on the spot orientation ($\chi$).

We now discuss the first step. The following steps were repeated for each of the 10201 diffraction patterns in a scan. The row, column and intensity vectors were extracted from the sparse matrix representing the ROI described above in a single diffraction pattern. Rows and columns were remapped (see equations above) into $q$ and $\chi$ coordinates. Row and column vectors defining the pixel positions in ROI of the diffraction pattern were clustered based on the pairwise Euclidean distance between the pixels.[78] We used a threshold of 0.8 for cutting the hierarchical tree.[78] As an example, the ROI of the first diffraction pattern of the scan displayed only one diffraction spot, i.e. only one cluster of 83 pixels. This is shown in Figure 14a in pixel coordinates. The cluster represented as a set of $q$, $\chi$ and intensity coordinates (inset in Figure 14a) was fitted with a cubic spline interpolant using a mesh with a regular $q$, $\chi$ step size (Figure 14b). The maximum intensity was extracted and the corresponding $q$, $\chi$ were labelled as the peak position. A line profile (Figure 14c) passing by the peak position and the origin of the diffraction pattern was extracted and used for the extraction of the left hand side width at half maximum (LHS), right hand width at half maximum (RHS), and the full width at half maximum (FWHM = LHS + RHS). This was the simplest case. Since the ROI used for this analysis includes two reflections (002) and (110), we also observed diffraction spots with double peaks. Figure 14d shows a diffraction spot displaying both reflections in the ROI of the third diffraction pattern. The most complicated case was when the ROI of a single diffraction pattern displayed multiple diffraction spots. This is shown in Figure 14g (image index 1718). The three diffraction spots were clustered as shown in the inset. Figure 14h and i show the previously described graphs for the second cluster. Once this procedure was iterated for the 10201 diffraction patterns, we generated a quiver[78] (or velocity) plot representing the $\chi$ coordinate of each detected peak in spatial coordinates. As for the montage, the first diffraction pattern, which displayed only one diffraction spot, resulted in an arrow with origin at (0.2, 0.2) μm, an angle corresponding to $\chi$, and a colour corresponding the a one to one correspondence between the angles from 0° to 360° and a jet colour map.[78] The above mentioned ROI of image index 1718, which was made of 3 diffraction spots in the ROI, resulted in three arrows with angles equal to 310.1°, 119.21°, 299.81°. A quiver plot was employed as it allowed visualizing multiple peaks on the same spatial position. This is illustrated in Figure S15.

We now discuss the second step. We performed as second level of clustering on $(x, y, \chi)$, where $\chi$ is the $\chi$ coordinate of diffraction spot and $x, y$ are the spatial coordinates of the point where the diffraction pattern containing the diffraction spot was taken. The purpose of the second level of clustering was to identify crystallites or grains based on their crystallographic orientation. The method used for this clustering is similar to the previous, with the difference that this one is performed in three dimensions and on different coordinates. Figure S16 shows results for the first two largest detected clusters, along with maps of the $q$, $\chi$ coordinates of the peak extracted from the diffraction spots within the clustered grain. In Figure S16b and c we can observe two regions corresponding to different $q$ and $\chi$ values. The region on top clearly belongs to the (002) reflection, while the one on the bottom belongs to the (110) reflection. An accurate analysis should take into account that (002) and (110) are different peaks. Peaks segregation had to be implemented in simple and robust way. We could not perform this step by defining a finer ROI separating the (002) from (110) reflections because the ROI was defined on the raw pattern, which presented distortion due to detector misalignment (non-circular pattern). Figure S17 shows a plot of $q$ vs. $\chi$ for the peak extracted from all the diffraction spots in the raster scan. These values were obtained after applying the distortion correction equations described above. The expected $q$ vs. $\chi$ curve should have been flat with some fluctuations. However, we found that the graph was best fitted by a 4$^{th}$ order polynomial (red curve), which confirmed that the distortion caused by the rotated detector was not completely removed. This was so far the best distortion correction we could obtain. We observed that the fitting curve could be used to discriminate between the (002) and the (110) reflections. Figure S18 shows the diffraction spot extracted from the 113$^{th}$ diffraction pattern (cold sample). The green line in Figure S18a is a plot of the 4$^{th}$ degree polynomial (shifted up for clarity), which passes between the two diffraction peaks. If we opt for the detection of the peak below this line, i.e. (002), we find $q$ = 0.9915 Å$^{-1}$ and $\chi$ = 228.14° as the detected peak. The line profile passing by $\chi$ = 228.14° is shown in Figure S18b. If we opt for the detection of peaks above this line, i.e. (110), we find $q$ = 1.0003 Å$^{-1}$ and $\chi$ = 227.51° as the detected peak. The line profile passing by $\chi$ = 227.51° is shown in Figure S18c. For both line profiles we fit the detected peak with a double Gaussian. Double Gaussians were used to take into account peak asymmetries, approximate quite well pseudo-Voigt profiles, and are easier and more robust to fit than pseudo-Voigt functions. Finally, we repeated the steps illustrated here twice for the ROIs corresponding to the 10201 images in a scan. The first time we focused on the (002) reflection, and the second time we focused on the (110) reflection.

### 4.6 Fabrication of Solar Cells

Glass substrates with 100 nm thick pre-patterned indium-tin-oxide (ITO) were obtained from Ossila Ltd (code S171, 6 pixels, 20 Ω/square). Substrates were cleaned using the following steps: (i) sonicate for 5 min in hot DI (de-ionized) water (~70°C) with ~1% Hellmanex III (Ossila, code C141); (ii) rinse twice in boiling DI water; (iii) sonicate for 5 min in isopropyl alcohol; (iiv) rinse in boiling DI water. Cleaned substrates were dried using compressed nitrogen and spin-coated with PEDOT: PSS. The steps followed for PEDOT:PSS and perovskite layers deposition in the preparation of the solar cells are identical to the steps detailed above for the preparation of the samples for SEM, GI-WAXS, and nXRD, thus achieving identical layer thicknesses. The substrates coated with the perovskite layer were annealed for 90 min at 90°C and transferred into a glovebox filled with nitrogen. [6,6]-Phenyl-C71-butyric acid methyl ester (PC$_{70}$BM, 99%, Ossila Ltd, code M113) was dissolved in chlorobenzene (50 mg ml$^{-1}$) at ~70°C and stirred for 12 hours (in the glovebox). The PC$_{70}$BM solution was then filtered using a 0.45 μm PTFE filter and spin-coated (1000 rpm, 30 s) on the perovskite substrates to obtain a 120 nm thick layer. Devices were finally introduced in a thermal evaporator and evaporated (base pressure ~1 × 10$^{-6}$ mbar) with 5 nm Ca and 100 nm Al using a cathode metal mask. After the cathode was evaporated, the devices were encapsulated with a UV-curable epoxy (Ossila, code E131) and thin glass cover slips.

## Supporting Information

https://www.dropbox.com/s/wmybufzc2mhlwi4/Supporting%20Information.docx?dl=0

## Acknowledgements

This work was funded by the UK Engineering and Physical Sciences Research Council via grants EP/M025020/1 'High resolution mapping of performance and degradation mechanisms in printable photovoltaic devices', EP/J017361/1 (Supersolar Solar Energy Hub) and the E-Futures Doctoral Training Centre in Interdisciplinary Energy Research EP/G037477/1. This work was partially funded by *The President of the UAE's Distinguished Student Scholarship Program (DSS), granted by the Ministry of Presidential Affairs, UAE* (M.A. PhD scholarship). This work was also partially funded by Masdar Institute through the grant *Novel Organic Optoelectronic Devices*. XMaS is a mid-range facility supported by the Engineering and Physical Sciences Research Council (EPSRC). We are grateful to all the beam line team staff for their support. We thank Ossila Ltd. for synthetizing methylammonium iodide and S. Saylan for the help with GI-WAXS measurements.